\newif\ifdraft
\drafttrue

\newif\ifresubmission
\resubmissionfalse

\documentclass[
twocolumn,
superscriptaddress,
aps,
prl,
]{revtex4-2}

\usepackage[T1]{fontenc}
\usepackage[utf8]{inputenc}
\usepackage[english]{babel}
\addto\captionsenglish{}
\addto\captionsenglish{}
\usepackage{graphicx}
\usepackage[colorlinks,breaklinks]{hyperref}
\usepackage{mathtools}
\usepackage{bm}
\usepackage{amssymb}
\usepackage{overpic}
\usepackage[section]{placeins}
\usepackage{enumitem}
\usepackage{microtype}

\makeatletter
\g@addto@macro\@floatboxreset{\centering}
\makeatother

\graphicspath{{figures/}}

\usepackage[normalem]{ulem}
\usepackage{xcolor}
\ifdraft

    \newcommand{\crossout}[1]{{\color{red}\sout{#1}}}
    \newcommand{\note}[1]{{\color{red} \textbf{[#1]}}}
 \else

    \newcommand{\crossout}[1]{}
    \newcommand{\note}[1]{}
 \fi

\begin{document}

\title{Low inertia limit of elasto-inertial turbulence }

\author{Shoaib Kamil}
\affiliation{Institute of Science and Technology Austria (ISTA), 3400 Klosterneuburg, Austria}
\author{Sarath Sankar Suresh}
\affiliation{Institute of Science and Technology Austria (ISTA), 3400 Klosterneuburg, Austria}
\author{Jose M. Lopez}
\affiliation{University of Malaga, Malaga, Spain}

\author{Björn Hof}
\email{bhof@ist.ac.at}
\affiliation{Institute of Science and Technology Austria (ISTA), 3400 Klosterneuburg, Austria}

\date{June 7, 2023}

\begin{abstract}

Pipe and channel flows of viscoelastic fluids display chaotic dynamics at unusually low speeds, a phenomenon referred to as elasto-inertial turbulence, EIT. First reported in experiments a century ago, recent theoretical studies and model computations predict a variety of scenarios for the phenomenon's origin, ranging from hoop stress modes to center modes and to Tollmien-Schlichting waves. Lacking experimental confirmation, the relevant scenario in actual flows of polymer solutions remains unknown. We here determine the transition threshold of EIT in pipe experiments, covering three decades in elasticity number. Across this entire parameter range, the transition features center mode structures at onset. Eventually the instability diverges at a lower inertia (upper elasticity) limit, which is a robust signature of this center mode scenario. Finally, we report the first experimental observation of a traveling wave in viscoelastic pipe flow, and the sequences of localized structures found, are in excellent agreement with a center mode traveling wave, the "arrowhead" solution, discovered in model simulations.

\end{abstract}
\maketitle

The addition of minute quantities of long chain polymers to a Newtonian solvent can fundamentally alter the motion of fluids. At high Reynolds numbers ($Re$), polymers dramatically decrease the drag of turbulent flows \cite{White-2008-ARFM}, a property frequently exploited to reduce pumping costs in oil pipelines. In contrast to this stabilizing effect, polymers cause instabilities \cite{larson1990purely, Shaqfeh-1996-ANRFM, mckinley1996rheological} and novel states of turbulence \cite{Groisman-2000-Nature, Groisman-2001-Nature, Samanta-2013-PNAS, yamani2021spectral} at low Re, a regime in which Newtonian flows are perfectly stable and laminar. Even though such low Re turbulent motions have already been reported for pipe flow a century ago \cite{Ostwald-1926-KOLL, Reiner-1926-1-KOLL}, their origin remains heavily debated \cite{dubief2023elasto}. Various studies proposed a link to a hoop stress instability that occurs in flows with curved streamlines \cite{larson1990purely} and here leads to purely elastic turbulence \cite{Groisman-2000-Nature, Steinberg-2021-ARFM}. While this instability cannot arise linearly in straight pipes and channels, it has been suggested to arise nonlinearly, i.e. following from perturbations of finite amplitude \cite{bertola2003experimental, Morozov-2005-PRL, Pan-2013-PRL, Morozov-2019-JSP}.

\begin{figure*}
  \includegraphics[width=0.8\linewidth]{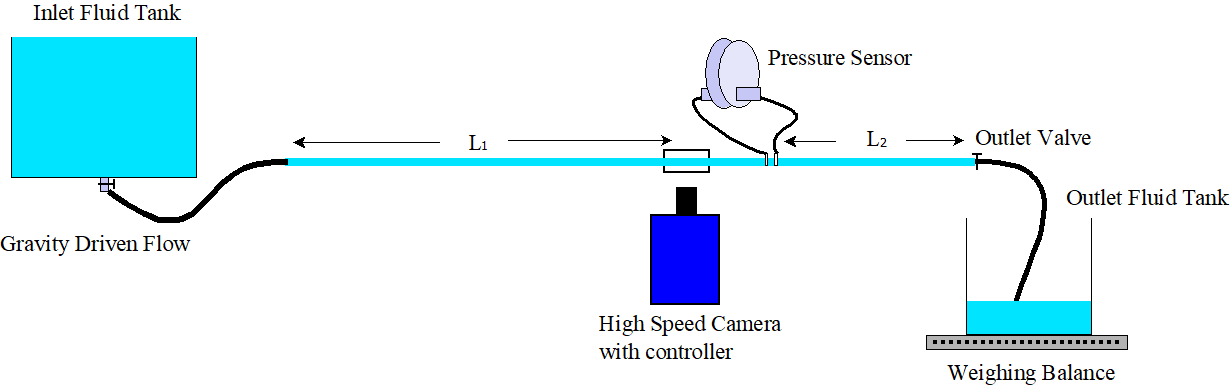}
\caption{Schematic of the experimental setup. The flow through a 4mm diameter long pipe is gravity driven. Measurements of pressure fluctuations and the velocity fields (by planar particle image velocimetry) were carried out $L_{1} = 175D$ downstream of the entrance, and additional tests were performed at $L_{1} = 275 D$ to verify that the results were independent of the measurement location. The exit length is both cases is $125 D$. The flow rate was determined by measuring the weight of the fluid collected in the outlet tank. }
   \label{fig:polymersetup_4mm}
\end{figure*}

More recently, a different linear instability mechanism \cite{Garg-2018-PRL, khalid2021centre} has been discovered for viscoelastic pipe and channel flows, i.e. a mechanism that does not require curved streamlines and is not driven by hoop stresses. Instead, it corresponds to a center mode instability and involves a balance between inertia, elastic stresses and solvent viscous effects \cite{Garg-2018-PRL, chaudhary2021linear}. For an increasing elasticity number, $E = 8\lambda\nu/D^2$, where $\lambda$ is the polymer relaxation time and $\nu$ the fluid's kinematic viscosity, linear stability analyses based on the Oldroyd-B constitutive model found that the critical Re decreases following a power law. However, for sufficiently large elasticity, the instability ceases to exist, setting a minimum $Re\approx63$ for pipe flow, with $Re=UD/\nu$ (where U is the bulk velocity, D the pipe diameter and $\nu$ the fluid's kinematic viscosity). 

As indicated by recent studies for channel flow this lower Re limit depends on the constitutive model\cite{Buza-2022-JFM, khalid2025role} and hence on fluid properties. Irrespective of the exact quantitative value, the occurrence of such an upper elasticity number bound is specific to the center mode instability and conversely does not arise for the competing scenario of the aforementioned purely elastic hoop stress instability \cite{larson1990purely}. 
On the one hand, the persistence of EIT in pipe experiments to elasticity numbers \cite{Choueiri-2021-PNAS} that are an order of magnitude larger than those of the linear center mode instability may be regarded as support of such a purely elastic scenario. On the other hand, the flow structures that were observed in these same experiments at onset have been found to resemble the center mode \cite{Choueiri-2021-PNAS}. 
Channel experiments \cite{Pan-2013-PRL} reported chaotic flow even at vanishing inertia ($Re<10^{-2}$) and elasticity numbers $O(10^3)$, a finding that has been interpreted in support of a subcritical purely elastic hoop stress transition route. 

An alternative explanation for the high elasticity numbers and low Re observed is that the center mode instability at this point occurs subcritically, in line with simulations of two dimensional channel flow \cite{page2020exact, dubief2022first, Wan-2021-JFM, Buza-2022-JFM}. This subcritical branch arises from a nonlinear traveling wave, referred to as the "arrowhead".  Subsequently, the arrowhead could be continued to negligible inertia \cite{morozov2022coherent} where it supports a purely elastic form of turbulence \cite{lellep2024purely}. 

Finally, other studies \cite{Shekar-2019-PRL, Shekar-2021-PRF} put forward a fundamentally different transition scenario, suggesting that EIT does not arise from elastic modes but essentially from a Newtonian solution branch, i.e. Tollmien-Schlichting waves. Computations in minimal channel flow units \cite{zhang2024minimal} recently argued in favor of this latter scenario based on observations of wall modes for sufficiently large domain sizes, whereas arrowhead structures (favoring a subcritical center mode route) were only found for the smallest domain sizes. While wall modes are also found to dominate the dynamics of actual EIT in experiments \cite{Choueiri-2021-PNAS}, in this case, the high amplitude near wall structures were found to only arise significantly above onset, whereas close to onset flows were dominated by the aforementioned center mode structures. 

As we will show in the following, in pipe flow experiments, the onset of EIT follows a simple power law scaling with the effective elasticity parameter E(1-$\beta$), where $\beta$ denotes the ratio of the solvent to the total viscosity. This scaling persists over a wide range, but ultimately the transition threshold diverges and the instability ceases to exist. This low Re, high E limit distinguishes the instability from a purely elastic (subcritical) hoop stress instability \cite{bertola2003experimental, Morozov-2005-PRL, Pan-2013-PRL, Morozov-2019-JSP}. Instead,
the scaling and the divergence of $Re_c$ qualitatively agree with the general shape of the neutral stability curve of the center mode instability \cite{Garg-2018-PRL}. The order of magnitude lower Re, and two orders of magnitude higher E values, however point towards the relevance of fluid properties neglected by present theory. Finally, the measured average flow profiles confirm the flows' center mode structure, while instantaneous velocity fields are in surprisingly close agreement with the arrowhead, and hence with a nonlinear center mode travelling wave.


\begin{figure*}
  \includegraphics[width=0.9\linewidth]{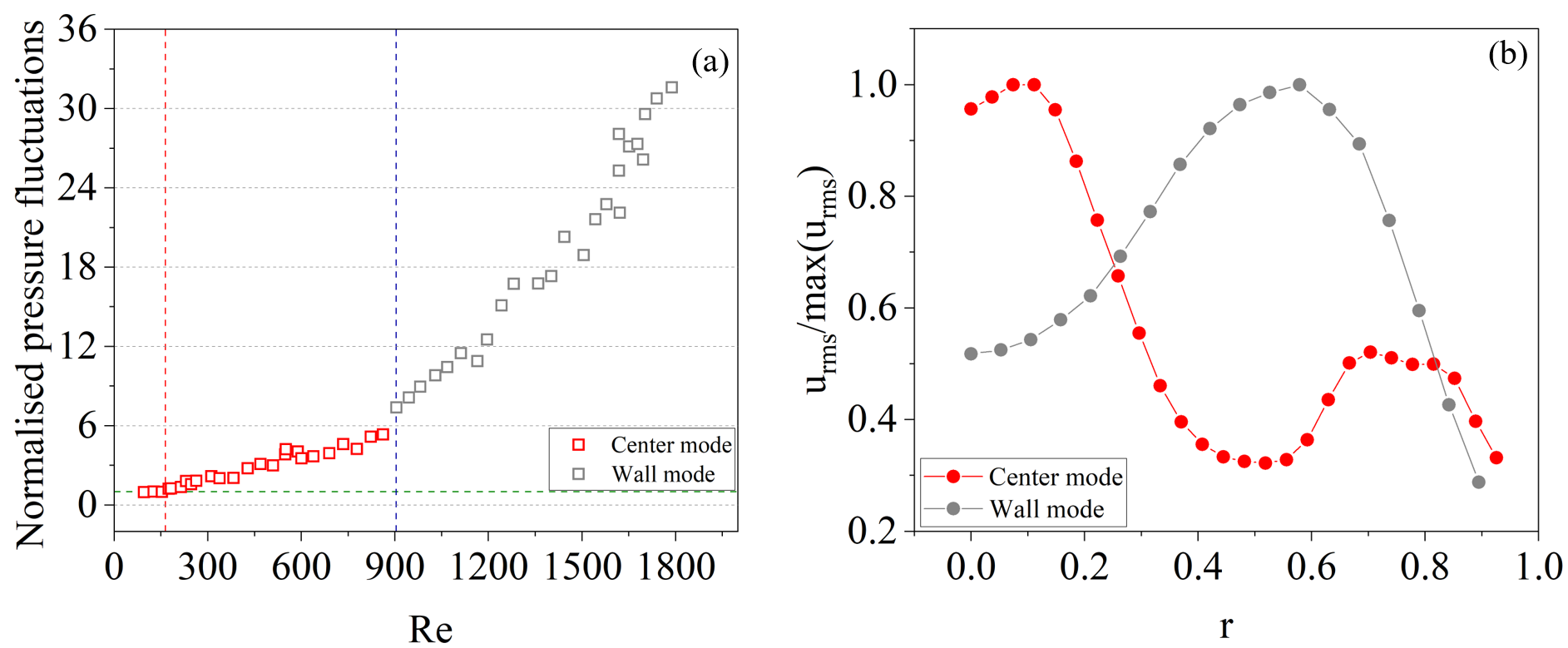}
  \caption{Pressure fluctuations demarcate the onset of EIT. (a) pressure standard deviations for 50 ppm PAAm dissolved in water.  The data is normalised by the respective sensor noise level in laminar flow. While the fluctuation level is constant in laminar flow ($Re < 150$), subsequent to the onset of EIT (red dashed line), pressure fluctuations increase gradually and in this initial range (red), the $u_{rms}$ profile corresponds to a center mode (red circles in (b)). From $Re\approx900$ (blue dashed line), the flow structure changes to a wall mode (gray circles in (b)) and the fluctuation amplitudes increase at a considerably larger rate (gray squares in (a)).}
   \label{fig:pressure fluctuation}
\end{figure*}

Experiments were carried out in a $D=4mm \pm 0.01$ diameter precision bore glass pipe, 300 D in length. To determine the deviation from laminar, we monitored pressure fluctuations using a differential pressure sensor (Validyne DP45) as illustrated in Fig~\ref{fig:polymersetup_4mm}. Moreover, we recorded velocity fields in the mid-plane (streamwise-radial) of the pipe using a high speed particle image velocimetry, PIV, system (LaVision GmbH). The flowrate was monitored using a precision balance (321 LX) and the temperature was recorded using a PT 100 probe (not shown).  Tests were performed with different inlet conditions (smooth versus abrupt diameter reduction) and with and without downstream perturbations. Moreover, the measurement point, typically 175D from the inlet, was shifted to 275D to ensure turbulence was fully developed. Like in earlier experimental studies of EIT\cite{Samanta-2013-PNAS}, fluctuation levels and, in particular, the detected transition point for the onset of EIT were found to be unaffected by these changes. Finally, we conducted a second set of experiments in a $D=7mm  \pm 0.01$ pipe setup to test the robustness of the phenomena found. 

In all experiments, the working fluid was a 50ppm solution of polyacrylamide PAAm (18 million Da) in water-glycerol mixtures of varying viscosity. Altering the solvent viscosity allows us to change E between 0.04 and 450. It is noteworthy that both, the value of E and the value of the viscosity ratio $\beta$ vary with shear rate. 
This property of polymer solutions constitutes a key difference to theoretical stability analyses based on the Oldroyd-B model \cite{Garg-2018-PRL}, which does not take shear thinning into account. In our experiments, the fluid's relaxation time and the shear dependent viscosity ratio were determined using an Anton Paar rheometer (MCR 102).Fluids were characterized before and after the measurement and no change in these properties, and hence no degradation could be detected. 
For the PIV measurements, the fluid was additionally seeded with a small amount of 3 micron sized silicon carbide particles.

Flows were found to be laminar at sufficiently low speeds (or low shear rates). In this steady flow regime, pressure fluctuations were indistinguishable from the sensor's intrinsic noise level. Upon increasing Re, the pressure fluctuations would eventually exceed this baseline level at a critical threshold. A representative example is shown in Fig~\ref{fig:pressure fluctuation}(a). Surpassing the critical point (for the example shown, $Re_c\approx150$) the fluctuation levels increases gradually and throughout this parameter range velocity profiles (measured using PIV) are characterized by a fluctuation peak close to the pipe center (red data set in Fig~\ref{fig:pressure fluctuation}(b)). Eventually at a sufficiently larger $Re$ and in agreement with the experiments of \cite{Choueiri-2021-PNAS}, the fluctuation level increases more rapidly (gray circles) and the flow structure transitions to a wall mode (gray symbols in Fig~\ref{fig:pressure fluctuation}(b)). This same sequence is found for all elasticity numbers investigated and the wall mode hence only arises far above the onset of EIT. Given that the present study is concerned with the transition from laminar flow, we solely focus on the primary instability branch and its scaling with elasticity number and the coherent flow structures found prior to the onset of the wall mode.

\begin{figure}
    \includegraphics[width=0.99\linewidth]{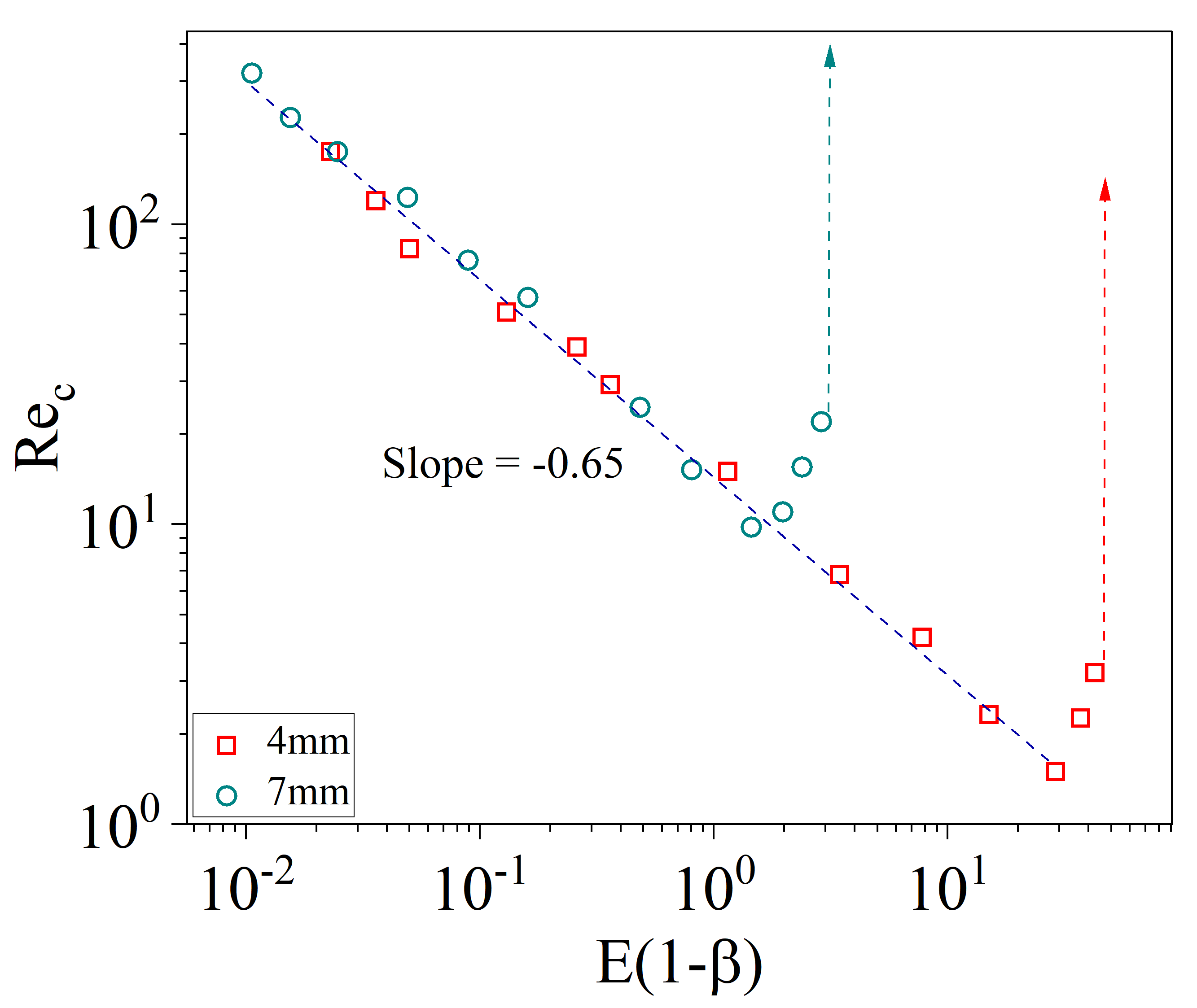}
    \caption{Transition threshold ($Re_c$) for the onset of EIT as a function of effective elasticity parameter ($E(1-\beta)$), measured in the $D=4 mm$ and $7 mm$ pipes. In both cases, $Re_c$ decreases with $E(1-\beta)$ following a power law with an exponent of $-0.65$. The instability threshold diverges at an upper elasticity number limit, which is a characteristic feature of the center mode instability \cite{Garg-2018-PRL}. In the larger diameter tube, the divergence occurs at larger $E(1-\beta)$, whereas in the smaller diameter tube, where shear thinning is more pronounced, it persists to elasticity numbers an order of magnitude larger. The destabilizing effect of shear thinning and the persistence of the center mode to lower Re and larger elasticity numbers qualitatively confirms recent stability analyses for constitutive models that incorporate shear thinning and finite extensibility \cite{Buza-2022-JFM, khalid2025role, Shankar, Morozov}.}
    \label{fig:4mm_50ppm}
\end{figure}

By changing the solvent viscosity, we trace the transition onset across three decades in $E(1-\beta)$ and, as shown in Fig.~\ref{fig:4mm_50ppm}, in this range the transition Reynolds number is found to decrease with a power law exponent of $-0.65$ (i.e. close to -2/3). This value is comparable to that found in earlier experiments \cite{Chandra-2020-JFM}. However, data from these earlier experiments showed much larger scatter and reported a dependence of the transition threshold on pipe diameter, whereas we find excellent data collapse and the same scaling exponent for the different pipe diameters investigated. 

The exponent of -0.65 considerably deviates from the value of $-3/2$ \cite{Garg-2018-PRL} predicted by the linear stability analysis of an Oldroyd-B fluid and hence a model lacking shear thinning and finite extensibility. Based on a simple scaling argument, these latter fluid properties have been suggested to change the threshold scaling in pipe flow to a value of $-0.625$ \cite{chaudhary2021linear}, consistent with the value found in the present experiments. It is noteworthy that, even for Oldroyd-B fluids in the intermediate elasticity number regime, exponents can strongly deviate from the asymptotic scaling and were found to be close to $-0.5$ \cite{khalid2021centre}. 


The exact value of the exponent aside, the most striking feature of the threshold scaling is a slope reversal and eventual divergence, a qualitative feature that is specific to the center mode \cite{Garg-2018-PRL}. However, in experiments, the instability extends to Re considerably lower than the minimum value predicted by the Oldroyd-B model ($Re\approx63$ \cite{Chaudhary-2021-JFM}). Moreover, while for small $E(1-\beta)$ the transition thresholds between the two different pipe experiments collapse, the divergence point differs. In the 7mm pipe, the minimum $Re_c\approx10$, in the 4mm pipe, the instability persists to a value of $Re_c\approx1$. A difference between measurements in pipes of different diameters may at first sight appear surprising, however at identical Re the shear rates in the two tubes differ by a factor of $\approx3$. Increasing shear decreases the fluid's viscosity due to shear thinning and with it $\beta$ increases (all $\beta$ values in Fig.~\ref{fig:4mm_50ppm} are based on the wall shear stress at the given $Re$). 

As shown recently in simulations of a FENE-P fluid in channel flow, shear thinning and finite extensibility lead to the persistence of the center mode instability to significantly lower Re (larger $E$)\cite{Buza-2022-JFM, khalid2025role}. Ongoing stability analyses for pipe flow, respectively using FENE-P and sPTT fluids, find the same trend and the center mode instability extends to Reynolds numbers of order unity\cite{Shankar, Morozov} and hence to the level predicted in the present experiments. The instabilities persistence to lower Re with increasing shear thinning also rationalizes the lower transition threshold observed in the 4mm pipe compared to the 7mm pipe, where shear thinning is less relevant.


In previous experiments \cite{Choueiri-2021-PNAS}, the flow structures close to the onset of EIT were found to resemble the center mode traveling wave but remained weakly irregular and time dependent. The low amplitudes of EIT close to onset combined with technical difficulties that arise from the small tube diameters, pose formidable challenges for spatially resolved velocity measurements. 

\begin{figure*}
  \includegraphics[width=1\linewidth]{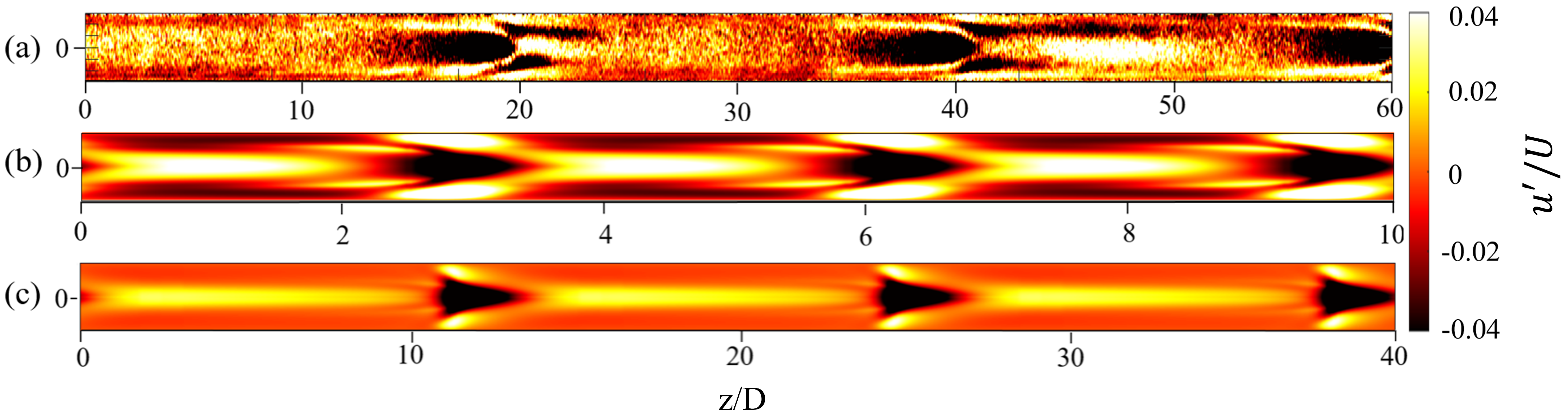}
  \caption{ (a) Coherent flow structures observed in the experiment at $E = 0.017$, $Re = 1600$. Equivalent coherent structures were observed close to the onset of EIT throughout the parameter regime. 
  In above case the coherent structures persist for 60 advective time units which in turn corresponds to 17 polymer relaxation times. The sequence closely resembles the arrowhead traveling wave recently found in channel flow simulations\cite{page2020exact}. For comparison we computed the arrowhead solution for pipe flow shown in panel (b) at $E = 1.2$, $Re = 25$ and in panel (c) $E = 1.2$, $Re = 100$. In both panels, the streamwise velocity fluctuations (with respect to the average velocity profile) are normalised with the respective bulk velocity. The experimental flow field is reconstructed by use of the Taylor frozen turbulence hypothesis.} 
  
   \label{fig:arrow_head}
\end{figure*}

Flow structures can be more readily resolved in pipes of larger diameter, since here the laser light sheet (typically 0.2mm thick) used to illuminate the pipe cross section, is better defined and less affected by tube curvature etc. We therefore initially switch to a $D=20mm$ pipe set up, that in all other aspects is similar to the set up described previously.  PIV measurements are carried out 175D from the inlet. We resolved the velocity field in the streamwise radial cross section as shown in Fig~\ref{fig:arrow_head}(a). Plotted are the deviations from the time-averaged mean profile, visualizing low- (black) and high- (red) speed streaks. Due to the larger tube diameter and hence lower shear rates, EIT is only found at somewhat larger Re, in this case the measurement was conducted at $Re=1600$ (see caption of Fig~\ref{fig:arrow_head} for flow parameters). The experimental measurement shows a sequence of three coherent bullet-shaped flow structures that in many aspects closely resemble the arrowhead traveling wave solutions previously reported for channel flow. For comparison, we computed the arrowhead for a two dimensional pipe flow using the FENE-P constitutive model (for more details about the code see \cite{Lopez-2019-JFM}). We show the solution for two sets of parameters E=1.2 and Re=25 in panel (b) and E=1.2 and Re=100 in panel (c) to highlight the changes in the structures appearance with parameters. Given that in addition to Re and E, shear thinning and the polydispersity of the  polymer molecules (resulting in a spectrum of relaxation times) introduce additional parameters, differences in the detailed appearance of the traveling wave in experiments are to be expected.

The degree of coherence observed in experiments, extending across twenty or so pipe diameters, is unprecedented when compared to turbulence in Newtonian pipe flow, where extensive searches only resulted in brief glimpses \cite{hof2004experimental,hof2008repeller, de2012edge} of structures that resemble those of nonlinear traveling waves. It is noteworthy that arrowhead structures are equally observable in the smaller diameter pipes and they persist throughout the Re-E parameter space investigated in our study. 


The persistence of coherent flow structures across many pipe diameters found in our measurements suggests that the dynamics of EIT are likely to be considerably lower dimensional than those of Newtonian pipe and channel flows, where flow fields are far more chaotic and only short glimpses of coherent structures have been reported \cite{hof2004experimental}.

    
It is noteworthy that the minimum Reynolds number in Fig~\ref{fig:4mm_50ppm} for the onset of EIT is fluid specific. Therefore, for higher polymer concentration or in smaller diameter tubes where shear thinning becomes more relevant, lower minimum thresholds and even the vanishing inertia limit can be reached in experiments as will be shown in a future study.


Overall, our observations demonstrate that the onset of EIT in experiments neither supports a hoop stress (purely elastic) scenario, nor a TS wave scenario. The former would imply a persistence of the instability with increasing fluid elasticity, while the latter would require wall-dominated modes at onset, whereas in our experiments, such wall modes only arise at substantially higher Re, well after the center-mode structure has appeared. All experimentally accessible signatures, the power-law scaling, the divergence at large elasticity, and the coherent structures at onset, identify the center-mode instability as the mechanism responsible for the onset of EIT in experiments.




\begin{acknowledgments}
The authors would like to thank Markus Holzner (BOKU, Vienna) for providing access to the $D=20mm$ pipe experiment. 

\end{acknowledgments}

\bibliographystyle{apsrev4-2}
\bibliography{bib}






\end{document}